\documentclass{aa}
\usepackage[varg]{txfonts}
\usepackage[utf8]{inputenc}
\usepackage{graphicx}
\usepackage{color}
\usepackage{natbib}
\bibpunct{(}{)}{;}{a}{}{,}
\newcommand{\cm}{cm$^{-1}$}

\begin{document} 

   \title{The FERRUM project: Experimental lifetimes and transition probabilities from highly excited even 4d levels in \ion{Fe}{ii}}
	\titlerunning{Lifetimes and $f$-values for 4d-levels in \ion{Fe}{ii}}
   \subtitle{}

   \author{          H. Hartman\inst{1,}\inst{2} \and  H. Nilsson\inst{2} \and L. Engstr\"om \inst{3} \and H. Lundberg\inst{3} }
   \authorrunning{Hartman, et al.}
         
   \institute{Material Sciences and Applied Mathematics, Malm\"o University, 20506 Malm\"o, Sweden\\
              \email{Henrik.Hartman@mah.se}\\
         \and
             Lund Observatory, Lund University, Box 43, 22100 Lund, Sweden \\
        \and
             Department of Physics, Lund University, Box 118, 22100 Lund, Sweden\\    
            }

   \date{Received ; accepted }
 
  \abstract{We report lifetime measurements of the 6 levels in the 3d$^6$($^5$D)4d e$^6$G term in \ion{Fe}{ii} at an energy of 10.4~eV, and $f$-values for 14 transitions from the investigated levels. The lifetimes were measured using time-resolved laser-induced fluorescence on ions in a laser-produced plasma. The high excitation energy, and the fact that the levels have the same parity as the the low-lying states directly populated in the plasma, necessitated the use of a two-photon excitation scheme. The probability for this process is greatly enhanced by the presence of the 3d$^6$($^5$D)4p z$^6$F levels at roughly half the energy difference. The $f$-values are obtained by combining the experimental lifetimes with branching fractions derived using relative intensities from a hollow cathode discharge lamp recorded with a Fourier transform spectrometer. 
The data is important for benchmarking atomic calculations of astrophysically important quantities and useful for spectroscopy of hot stars.
 }


\maketitle

\section{Introduction}
The present work reports on measurements of transition data involving high-excitation levels and lines of singly ionized iron, \ion{Fe}{ii}. 
The complex and line rich spectra of iron are observed in a wide variety of objects, such as stars, interstellar medium, solar coronae and quasars. 
In spectroscopy of stellar photospheres the atomic transitions most often appear as absorption lines. To use the spectral line for quantitative analysis of the population distribution and abundance determination in a stellar object, reliable $f$ values must be available. 

To meet the demands for accurate atomic data, \citet{JDD02v2} initiated the FERRUM project with the goal of providing evaluated transition data for astrophysical applications. Transitions from levels with different excitation potentials, and with a range of transition probabilities have been measured and calculated for the iron group elements, see e.g. \citet{ssk99,HSL05,GNE10}, and references therein. The present paper extends the data of highly excited levels, around 10 eV. The amount of data available from experimental studies is limited, and the majority must be provided by calculations. The experimental data is important to benchmark different calculations and to provide the user with reliable uncertainties and priorities between data sets. This is especially true for the complex ions where level mixing is important and hard to predict.

A common technique to derive the oscillator strength is to combine measurements of the radiative lifetime of an upper level with relative intensities for all decay channels. The present paper reports on values using this approach, where the radiative lifetimes are determined with Time-Resolved Laser Induced Fluorescence (TR-LIF) from ions produced in a laser ablation plasma. These are combined with branching fractions obtained from intensity calibrated Fourier transform spectra using a hollow cathode discharge as a light source. 

\section{Radiative lifetime measurements}

\begin{figure}[tbhp]
\begin{center}
\includegraphics[height=120mm]{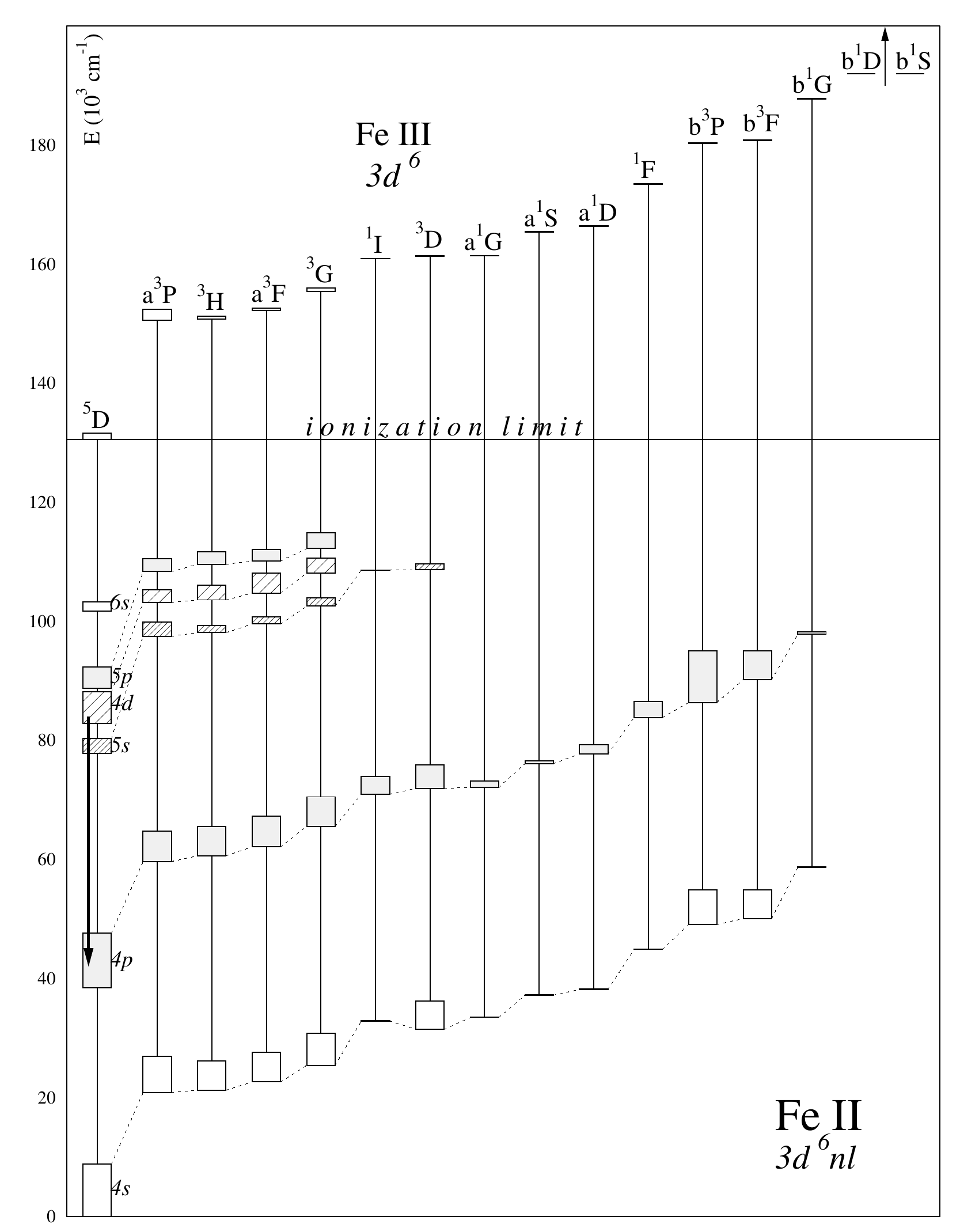}
\caption{A partial level diagram of \ion{Fe}{ii} showing the parent term structure, and the similar energies for all 4s-4p and 4p-5s/4d transitions. The arrow marks the transitions investigated in this study.} 
\label{fig:energydiagram}
\end{center}
\end{figure}

Figure \ref{fig:energydiagram} presents a partial energy level diagram for \ion{Fe}{ii} showing the large number of parent terms. The close similarity in the energy differences between 4s and 4p for all parent terms (the promotion energy), give rise to a high \ion{Fe}{ii} line density in the region around 230-260 nm. In addition, the 4s-4p and 4p-5s/4d transition arrays have similar energy differences, and the latter transitions thus also appear in the same wavelength region.
The target of the present study is the 3d$^6$($^5$D)4d~e$^6$G term. This term has even parity which is the same as the ground configuration 3d$^6$($^5$D)4s. This, and the large energy difference makes e$^6$G impossible to excite in a one-photon process. However, the use of an intense laser beam opens the possibility for two-photon excitations from the ground term, which has a high relative population in the ablation plasma. 

\begin{figure}[tbhp]
\begin{center}
\includegraphics[width=90mm]{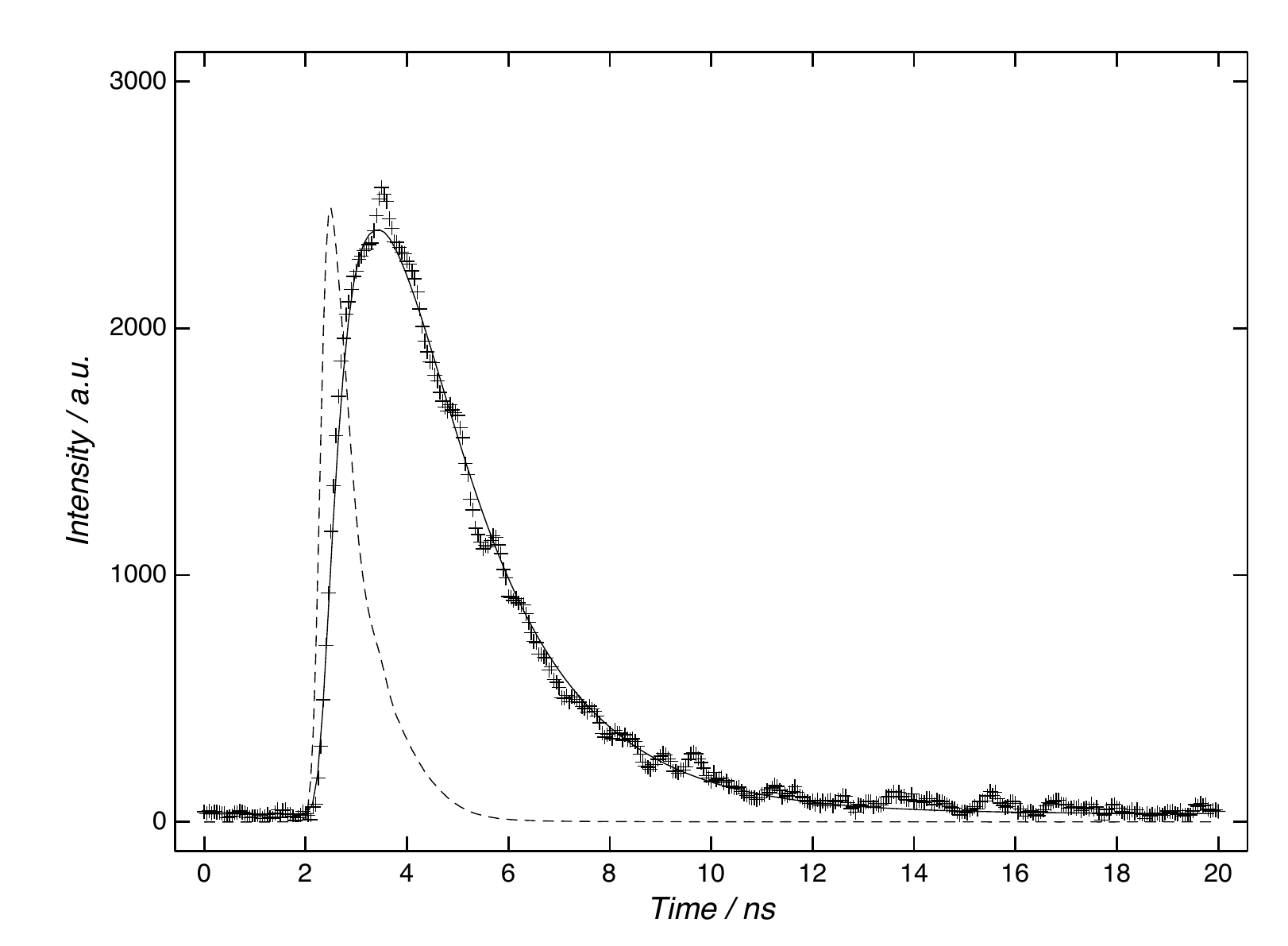}
\caption{The first 20 ns of the decay of the 3d$^6$4d e$^6$G$_{11/2}$ level in \ion{Fe}{ii} following two-photon excitation
from the ground state (3d$^6$4s a$^6$D$_{9/2}$). The evaluated lifetime is 2.0$\pm$0.2 ns. Background
subtracted data points (+) are plotted together with a fitted single exponential decay convoluted
by the square of the measured laser pulse (solid line). The dashed curve shows the squared
laser pulse. }
\label{fig:decaycurve}
\end{center}
\end{figure}

The TR-LIF set-up at Lund High Power Laser Facility is recently described in detail by \citet{ELN14}, where a study similar to this, but for \ion{Cr}{ii}, is reported; here we only emphasize the important aspects.  The set-up contains two Nd:YAG lasers operating at 10 Hz. One of them (Continuum Surelite) is frequency doubled and focussed onto a rotating iron target placed inside a vacuum chamber to generate the ablation plasma. The second laser is an injection seeded and Q-switched Continuum NY-82. The 532~nm output from this laser was temporally compressed using stimulated Brillouin scattering in water before pumping a dye laser (Continuum Nd-60) using DCM\footnote{4-(Dicyanomethylene)-2-methyl-6-(4-dimethylaminostyryl)-4H-pyran}. To obtain the 237~nm radiation needed for the two-photon excitation, the output was frequency tripled in KH$_2$PO$_4$ (KDP) and BaB$_2$O$_4$ (BBO) crystals and Raman shifted in a high pressure hydrogen cell.   
The laser-induced fluorescence was detected by a 1/8 m monochromator with a 280 $\mu$m wide entrance slit oriented parallel to the excitation laser beam and perpendicular to the ablation laser. All measurements were performed in the second spectral order giving an instrumental line width of 0.5~nm. The fluorescence signal was recorded with a micro-channel plate photomultiplier (PM) tube (Hamamatsu R3809U) with a rise time of 0.2~ns, and digitized by a Tektronix oscilloscope with 2.5~GHz analogue bandwidth. A second channel on the oscilloscope sampled the excitation laser pulse shape from a fast diode fed by scattered light from an attenuating filter. Each decay curve, as well as the temporal shape of the excitation pulse, was averaged over 1000 laser shots. The lifetimes were extracted using the program DECFIT \citep{PQF08}. The program derives the lifetime from the measured laser pulse and the recorded decay. 
In the present experiment, the fluorescence data was fitted with a single exponential convoluted with the square of the measured laser pulse (since the upper state is excited through a two-photon absorption) and a constant  background. The laser pulse had a temporal width (FWHM) of 2~ns. A typical example is shown in Figure~\ref{fig:decaycurve}. 

The final lifetimes given in Table \ref{tab:lifetimes} are the averages of around 20 measurements, performed during different days. The quoted uncertainties, between 10 and 17\%, include the statistical contributions from the fitting process as well as variations in the results between the different measurements.
The high probability of spectral coincidences between the two-photon pumping channel and a single photon excitation of the
4p levels had to be thoroughly considered. Another consequence of the similarity between the 4s-4p and 4p-4d wavelengths is that the secondary cascades (4s-4p) coincide with the measured primary decay (4p-4d). The chosen pump and detection channels are given in Table\ref{tab:lifetimes}, and the most difficult cases will be discussed in detail below. 
All measurements are checked for blending lines, both in the excitation and decay channels, and three of the levels are discussed in more detail below.

The high $J$ value of the e$^6$G$_{13/2}$ level, and the $\Delta J=0,\pm 1$ restriction for E1 transitions, effectively limits the decay channels to only the ($^5$D)4p level with $J$=11/2,  z$^6$F$_{11/2}$. Transitions to 4p levels belonging to other parent terms are several orders of magnitudes weaker. 
The pumping transition is located at 238.0~nm, and the primary and secondary decays fall at 237.64~nm and 238.28~nm, respectively. In this case the detection had to be shifted towards shorter wavelengths to avoid the influence from scattered laser light and secondary decay. 
However, shifting the detection wavelength too much to the blue brings in the fluorescence at 234.8~nm from 4p~z$^4$D$_{7/2}$ with a lifetime of 3.02~ns \citep{GAP92}, which is excited by a single 238.0~nm photon. This effect was clearly observed in the decay curve as a weak multi-exponential shape which resulted in a longer lifetime in our single exponential analysis compared to detection at longer wavelengths. 
By stepping the detection wavelength from this decay towards the primary decay at 237.64~nm, the contribution from z$^4$D$_{7/2}$ could be monitored, both in the derived lifetime and the fitted curve. From this behaviour and the known instrumental line width of 0.5~nm we estimated the maximum contribution from other decays to be 10\%.  This is reflected in the larger total uncertainty quoted for this level, 14\%, compared to 11\% for the other levels.

The level e$^6$G$_{11/2}$ was pumped at 237.2 nm from the ground state a$^6$D$_{9/2}$. The primary decay is at 237.0 nm whereas the secondary decay, a$^6$D-z$^6$F, falls at 238.2 nm. The monochromator was tuned to the blue side of the decay channel to minimize scattered laser light in the detection. In addition, the contribution from secondary decays was avoided. 

The level e$^6$G$_{7/2}$ can be populated by two-photon excitation from the lower levels a$^6$D$_{3/2, 5/2, 7/2}$. The excitation from $J$=3/2 was the best channel. The transitions from $J$=7/2 is close to the excitation route to e$^6$G$_{13/2}$, but still possible. The transition from $J$=5/2 is close to the laser frequency and could not be used. 

\begin{table*}
\caption{Excitation and detection scheme for the measured levels, and results from this work and comparisons with previous calculations.} \label{tab:lifetimes}
\begin{tabular}{llllllll} \hline
Level &  $E$ / cm$^{-1}$ & \multicolumn{2}{l}{Pump channel} & Detection channel & \multicolumn{3}{l}{Lifetime / ns }\\
 &  & Level & $\lambda$ / nm & $\lambda$ / nm & This work & K13$^a$ & RU98$^b$ \\ \hline \hline
e$^6$G$_{13/2}$ & 84035 &  a$^6$D$_{9/2}$ & 237.99 & 237 & 2.1$\pm$ 0.3 & 1.48 & 1.57 \\ 
e$^6$G$_{11/2}$ & 84296  &  a$^6$D$_{9/2}$ & 237.26 & 236-237 & 2.0$\pm$ 0.2 & 1.46 & 1.56  \\
e$^6$G$_{9/2}$ & 84527 &  a$^6$D$_{7/2}$ & 237.69 &  235-238 & 1.9$\pm$ 0.2 & 1.45 & 1.55  \\
e$^6$G$_{7/2}$ & 84710 &  a$^6$D$_{3/2,7/2}$ & 237.10; 238.45 &  235-263& 1.8$\pm$ 0.2 & 1.55 & 1.55  \\
e$^6$G$_{5/2}$ & 84844 &  a$^6$D$_{5/2}$& 237.52 & 235-266 & 1.8$\pm$ 0.2 & 1.44 & 1.54 \\
e$^6$G$_{3/2}$ & 84938 &  a$^6$D$_{1/2}$ & 238.21 & 235-236 & 1.7$\pm$ 0.3 & 1.43 & 1.53 \\ \hline
\end{tabular}\\
$^a$\citet{K13},
$^b$\citet{RU98web,RU98}
\end{table*}

\begin{figure}[tbhp]
\begin{center}
\includegraphics[width=\columnwidth]{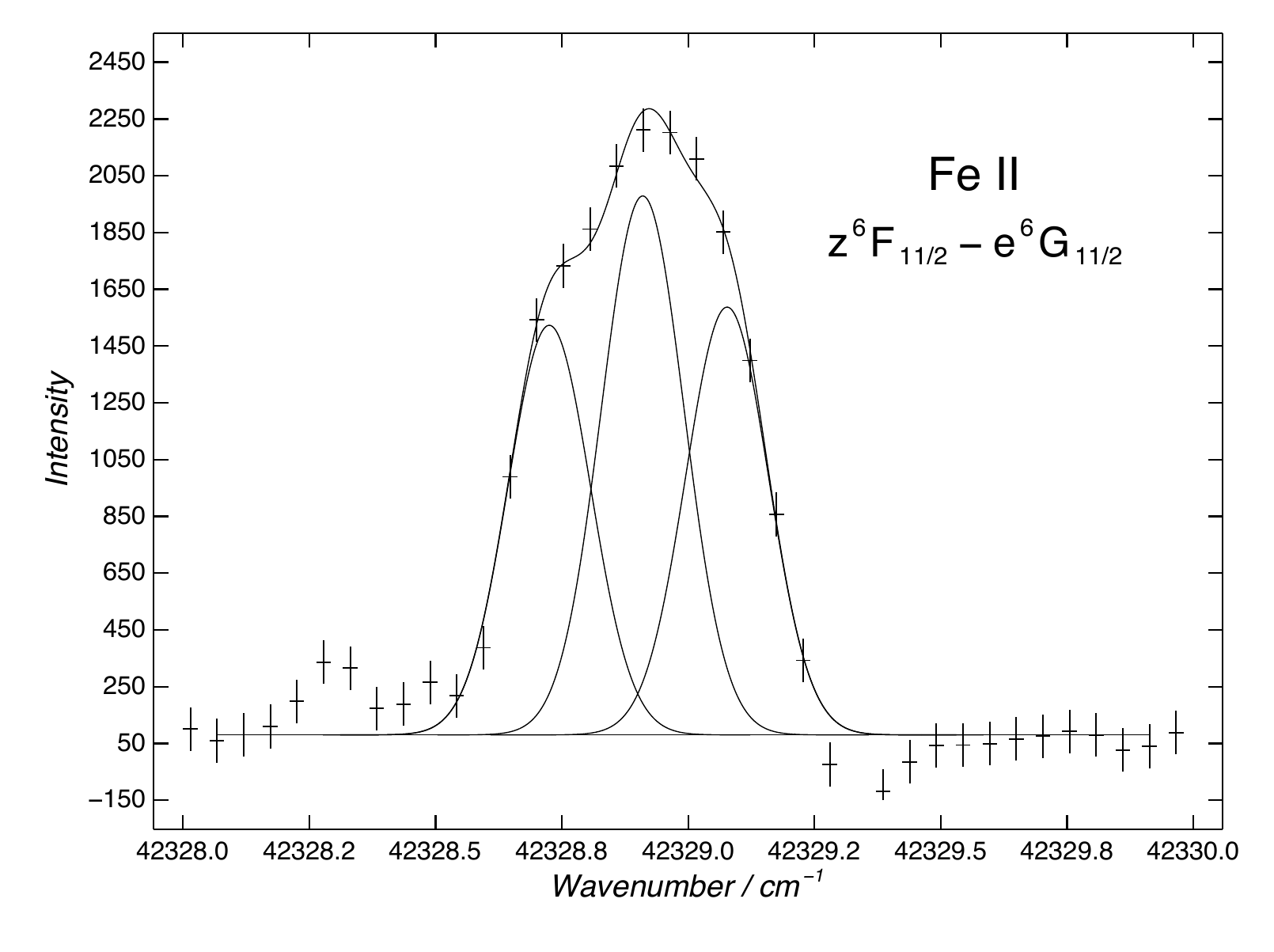}
\caption{The 4p z$^6$F$_{11/2}$ - 4d e$^6$G$_{11/2}$ line at 42328.78 \cm blended by two other \ion{Fe}{ii} lines at 0.185 and 0.352 \cm higher wavenumber, respectively. These separations have been kept fixed in the fitting of the three Gaussian shaped lines.}
\label{fig:ftsspectrum}
\end{center}
\end{figure}

\section{Branching fractions}

The \textit{BF} for a transition from an upper level $i$ to a lower level $k$ can be expressed as:

\begin{equation} 
BF_{ik} = \frac{A_{ik}}{\sum\limits_{j=1}^{N} A_{ij}} = \frac{I_{ik}}{\sum\limits_{j=1}^{N} I_{ij}} 
\label{eq:bf}
\end{equation} 

Here $N$ is the total number of transitions from level $i$ and $j$ is summed over all lower levels. The first equality is the definition of $BF_{ik}$, i.e. the fraction of the atoms in level $i$ that decay through the channel $ik$ . The second equality assumes the case of an optically thin plasma, where the observed photon intensity of a line, $I_{ij}$ measured as photons per second, is proportional to $A_{ij}$ \citep{TLJ99}  . When all intensities are measured on a common calibrated scale, $BF_{ik}$ can be derived from the measured intensities. 

The lifetime of the upper level, $\tau_i$, puts the relative intensities on an absolute scale. With 
\[ \tau_i = \frac{1}{\sum\limits_{j=1}^{N} A_{ij}} \]
 and Eq.~\eqref{eq:bf} we derive $A$ from 
\begin{equation} 
A_{ik} = \frac{BF_{ik}}{\tau_i} 
\label{eq:A}
\end{equation} 

\begin{figure}[htb]
\begin{center}
\includegraphics[width=\columnwidth]{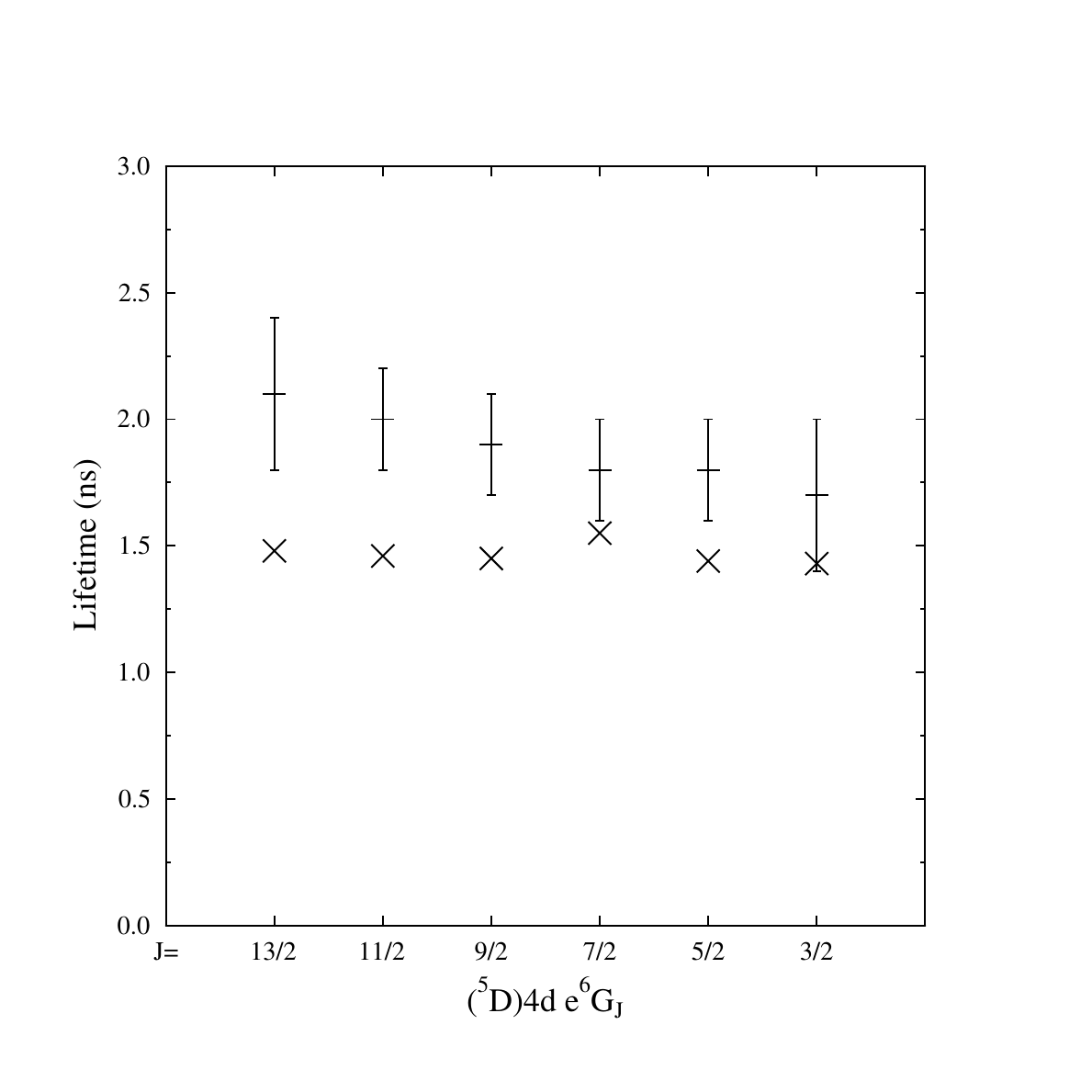}
\caption{Experimental lifetimes are plotted with error bars (+), whereas the theoretical lifetimes \citep{K13} are marked with crosses (x).}
\label{fig:lifetimes}
\end{center}
\end{figure}

This requires all lines from the upper level to be measured. In practice it is usually impossible to include all transitions, either because they occur outside the detector range or they are too weak to be observed. Thus, we rewrite Eq.~\eqref{eq:bf} as:

\begin{equation} 
BF_{ik} 
=  \frac{I_{ik}}{\sum\limits_{j=1}^{N} I_{ij}} 
= \frac{1}{1+ \sum\limits_{j=1, j \neq k}^{n} \frac{I_{ij}}{I_{ik}} + \sum\limits_{j=n+1}^{N} \frac{I_{ij}}{I_{ik}}} \approx \\
\label{eq:residual}
\end{equation} 
\[
\approx \frac{1}{1+ \frac{1}{I_{ik}} \sum\limits_{j=1, j \neq k}^{n} I_{ij} + \frac{1}{A_{ik}}\sum\limits_{j=n+1}^{N} A_{ij}}
\]

Here $n$ is the number of observed lines and the sum over the intensity ratios for the unobserved lines is approximated using the theoretical transition probabilities from \citet{K13}.

\begin{figure*}[tbhp]
\begin{center}
\includegraphics[width=\textwidth]{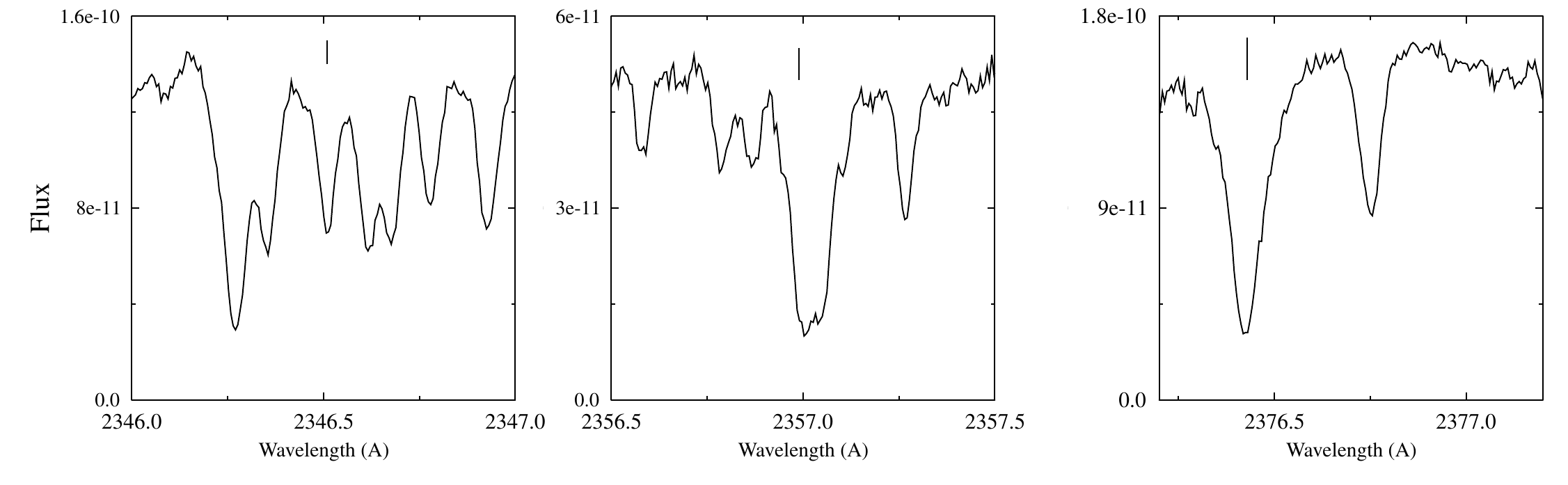}
\caption{Partial absorption spectrum of the chemically peculiar star chi Lupi, with the investigated \ion{Fe}{ii} lines marked with tickmarks. The spectrum is observed by the Hubble Space Telescope and its Goddard High Resolution Spectrograph, and is obtained from the MAST data archive at the Space Telescope Science Institute. See \citet{BHB99} for a complete identification list and a comparison with a synthetic spectrum.}
\label{fig:chilupi}
\end{center}
\end{figure*}

The iron plasma was produced in a hollow cathode discharge (HCD), operated with neon as carrier gas and using currents between 0.6 and 0.8 A and pressures between 1 and 2 torr. The light emitted was recorded with a Fourier transform spectrometer (Chelsea Instruments FT500) in the range 27640 - 52290 cm$^{-1}$ and detected by a solar blind PM tube (Hamamatsu R166). In total, five spectra were measured. While a higher current in the HCD produced a more intense spectrum it also led to slightly wider lines due to the temperature dependent Doppler effect. Line widths between 0.18 and 0.25 \cm were observed with most of the spectra having 0.21 \cm. To calibrate the intensity scale, three spectra with lower resolution were recorded using a deuterium lamp with known relative spectral radiance, measured at the Physicalisch-Technische Bundesanstalt, Berlin, Germany.

Despite the high spectral resolution of the spectrometer (resolving power about $2\cdot10^5$) the very line rich spectrum of Fe$^+$ in this region produced a number of blends that must be taken into account to derive accurate experimental intensities. The first case is the z$^6$F$_{11/2}$ - e$^6$G$_{11/2}$ channel at 42328.78~cm$^{-1}$, where there are two additional \ion{Fe}{ii} lines at 0.185 and 0.352~\cm higher wavenumbers, respectively \citep{J78,NJ13}. This blending was handled by fitting a sum of three Gaussian shaped line components, constrained to have the same FWHM and their separations fixed to the accurately known values, using the computer code GFit \citep{E14}.  One of these fits is shown in Figure \ref{fig:ftsspectrum}. Consistent line widths and relative intensities were obtained in all five measurements.

A more severe blending is found in the z$^6$F$_{5/2}$ - e$^6$G$_{7/2}$ channel at 42375.86~\cm where another \ion{Fe}{ii} line is only 0.11~\cm higher. Furthermore, contrary to the previous case the blending line is also significantly more intense. The experimental spectra were analysed using the same technique of constrained fitting as before. Also, in this case, we obtained the same FWHM as for resolved lines but the relative intensities of the two components varied somewhat between the five measurements. 
In addition to the three lines included in our work, \citet{K13} predicts an $A$ value larger than 10$^7$ s$^{-1}$ for four additional transitions (41472, 40478, 40263 and 39925 cm$^{-1}$) from the e$^6$G$_{7/2}$. Unfortunately all these lines are too severely blended in our spectra to be included in our analysis. This causes the so-called residual in Table 2, i.e. the sum of all theoretical $BF$s from \citet{K13} for the unobserved lines from this level, to be as high as 34\%. Since the experimental $BF$s, according to Eq (3), depends on this residual, our values for the e$^6$G$_{7/2}$ level must be considered less reliable than for the other levels which is reflected in the higher uncertainties.

The last example of blending is the close coincidence between z$^6$F$_{7/2}$ - e$^6$G$_{5/2}$ at 42607.80~\cm and a \ion{Fe}{ii} transition at 42607.79~\cm. In addition, the lines can be expected to have similar intensities. Thus the 7/2 - 5/2 combination are omitted from the $BF$ analysis. 

We present the final average $BF$s and transition probabilities obtained in Table \ref{tab:Avalues}, where we include the theoretical results by \citet{K13} for comparison. The error estimate in the $BF$s includes the uncertainty in the intensity calibration, in the area determination of the peaks, and a 50 \% uncertainty in the contribution from the unobserved branches obtained from \citet{K13}, as described in Eq~\eqref{eq:residual}. Since all but one of the observed decay branches are within 735 \cm of each other and, in addition, occur in a wavenumber region where the deuterium lamp calibration spectrum is quite flat, the uncertainty contribution from the calibration is small, about 3~\%. However, in the present case the total uncertainty of the transition probabilities for the strong lines is dominated by the contribution from the lifetime measurements. For the weak lines, the uncertainty in the $BF$ is the dominating source. 
A more general discussion of uncertainties in experimental transition probabilities can be found in \citet{SNL02}.

\section{Discussion}
Figure \ref{fig:chilupi} shows a \textit{Hubble Space Telescope} spectrum of the binary system chi Lupi (\object{HD141556}), which has spectral types B9.5p HgMn and A2 Vm with effective temperatures of $T_{\mathrm{eff}}$=10\,650~K and 9200~K, respectively \citep{WAR94}.
 The investigated transitions 4p--4d~e$^6$G are observed as prominent absorption features around 235~nm.

 In Table \ref{tab:lifetimes} we compare our experimental lifetimes with the available theoretical data from \citet{K13} and \citet{RU98web}. \citet{K13} applied a superposition of configurations method using a modified version of the Cowan codes \citep{C81}, whereas \citet{RU98}  applied the orthogonal operator formalism. Both calculations make use of experimental level energies to improve the results. As seen in Table \ref{tab:lifetimes}, the two theoretical results are within 7\%, but they are about 25 \% shorter and do not overlap with the experimental values, even taking the rather large uncertainties into account. A graphical comparison is shown in Figure \ref{fig:lifetimes}, where a weak tendency of lower experimental lifetimes for lower $J$ values is seen. This trend is within the error bars. 
 
The discrepancy between measured and calculated $BF$s is relatively small for most lines. An interesting exception are the transitions from e$^6$G$_{7/2}$, where there is a large difference between the two theoretical studies. Although both give similar lifetimes they differ significantly in the transition probabilities for individual lines. 
For two of the transitions our values support those of \citet{K13}, whereas the z$^4$P$_{5/2}$ -  e$^6$G$_{7/2}$ line at 37743 cm$^{-1}$, which is predicted to be quite intense is barely observable in our spectra. In addition, the large residual for this level, as discussed above, leads to a large uncertainty in the experimental $A$ values. 
The fact that the values by \citet{K13} differ significantly from his earlier 2010 calculations indicate theoretical difficulties for this level. \footnote{The calculations from 2010 (referenced as K10 in Kurucz notation) are not available online any longer, but replaced by the new data \citet{K13}.}. 
However, for all levels the resulting experimental $A$ values are a combination of lifetimes and $BF$s, and the discrepancy between the experimental and theoretical results is in most cases a consequence of the lifetime differences.
 
The general agreement, except for the mentioned e$^6$G$_{7/2}$, of the experimental $f$ values with the calculations by \citet{K13} and \citet{RU98web} provides us with confidence to use either set for abundance determinations and also for high-excitation lines represented by the transitions and levels investigated in this work. We have a preference for the calculations by \citet{K13} when compared with the experimental values, however, the discrepancy between experiment and theory has to be considered when estimating the uncertainty in the abundance determination or evaluation of the stellar atmosphere models. We recommend that the experimental data is used where available, and complemented with theoretical calculations for the other lines.
 

\begin{acknowledgements}
We are grateful to the anonymous referee for valuable comments significantly improving the manuscript. This work was supported by the Swedish Research Council through the Linnaeus grant to the Lund Laser Centre and the Knut and Alice Wallenberg Foundation. HH gratefully acknowledges the grant no 621-2011-4206 from the Swedish Research Council and support from the The Gyllenstierna Krapperup's Foundation. Figure \ref{fig:chilupi} is based on observations made with the NASA/ESA Hubble Space Telescope, obtained from the MAST data archive at the Space Telescope Science Institute. STScI is operated by the Association of Universities for Research in Astronomy, Inc. under NASA contract NAS 5-26555. 
 \end{acknowledgements}
\bibliographystyle{aa} 
\bibliography{../hartman} 

\begin{sidewaystable*}
\caption{The resulting transition probabilities and oscillator strengths for the lines studied.}
\begin{center}
\begin{tabular}{llllllllllrr} \hline
Upper level$^a$	& $\tau ^b$ / ns	& Lower level$^a$	& $\sigma^a$ / cm$^{-1}$	& $\lambda_{air}$ / nm & 	BF$_{exp}^b$	& BF$_{th}^c$	& A$_{exp}^b$ / 10$^8$ s & A$_{th}^{c}$ /10$^8$ s & A$_{th}^{d}$ /10$^8$ s & log \textit{gf} $^b$  & Uncert $^e$/ \% \\ \hline \hline
e$^6$G$_{13/2}	$ & 2.1$\pm$0.3 & z$^6$F$_{11/2}$ & 42067.10 & 237.64 & 1.00 & 1.00 & 4.76$\pm$0.7 & 6.32 & 6.36 & 0.752 & 15 \\[3mm]
e$^6$G$_{11/2}$ & 2.0$\pm$0.2 & z$^6$F$_{11/2}$ & 42328.81 & 236.17 & 0.0595$\pm$0.003 & 0.0666 & 0.298$\pm$0.03 & 0.455 & 0.434 & -0.524 & 10 \\
				& & z$^6$F$_{9/2}$ & 42182.04 & 237.00 & 0.928$\pm$0.007 & 0.921 & 4.64$\pm$0.5 & 6.30 & 5.88 & 0.672 & 11\\
				& & Residual$^f$ & & & & 0.0124 & &  & & \\[3mm]		
e$^6$G$_{9/2}$	& 1.9$\pm$0.2	& z$^6$F$_{9/2}$ & 42412.97 & 235.70 & 0.146$\pm$0.006 & 0.141 & 0.769$\pm$0.09 & 0.969 & 0.928 & -0.193 & 12\\
				& & z$^6$F$_{7/2}$ & 42290.75 & 236.39 & 0.826$\pm$ 0.01 & 0.832 & 4.34$\pm$0.5 & 5.72 & 5.31 & 0.561 & 12\\
				& & z$^6$P$_{7/2}$ & 41869.55 & 238.76 & 0.0137$\pm$0.005 & 0.0133 & 0.072$\pm$0.03 & 0.0916 & 0.106 & -1.210 & 42\\
				& & Residual & & & & 0.0145 & & & & \\[3mm]		
e$^6$G$_{7/2}$	& 1.8$\pm$0.2	& z$^6$F$_{7/2}$ & 42473.69 & 235.37 & 0.171$\pm$0.04 & 0.129 & 0.949$\pm$0.2 & 0.832 & 1.39 & -0.200 & 21\\
				& & z$^6$F$_{5/2}$	 & 42375.90 & 235.91 & 0.420$\pm$0.08 & 0.382 & 2.33$\pm$0.5 & 2.46 & 4.74 & 0.192 & 21\\
				& & z$^4$P$_{5/2}$	 & 37743.27 & 264.87 & 0.0070$\pm$0.003 & 0.144 & 0.039$\pm$0.02 & 0.926 & 0.0554 & -1.484 & 51\\   
				& & Residual & & & & 0.345 & & & & \\[3mm]		
e$^6$G$_{5/2}$	& 1.9$\pm$0.2	& z$^6$F$_{5/2}$ & 42510.02 & 235.17 & 0.276$\pm$0.01 & 0.274 & 1.45$\pm$0.2 & 1.91 &1.72 & -0.142 & 14\\
				& & z$^6$F$_{3/2}$ & 42443.55 & 235.54 & 0.692$\pm$0.01 & 0.693 & 3.64$\pm$0.4 & 4.82 & 4.42 & 0.259 & 11\\
				& & Residual & & & & 0.032 & & & & \\[3mm]		
e$^6$G$_{3/2}$	& 1.7$\pm$0.3  & z$^6$F$_{5/2}$ & 42603.41 & 234.65 & 0.0281$\pm$0.001 & 0.0267 & 0.165$\pm$0.03 & 0.186 & 0.178 & -1.264 & 18\\
				& & z$^6$F$_{3/2}$ & 42536.96 & 235.02 & 0.343$\pm$0.01 & 0.293 & 2.02$\pm$0.3 & 2.04 & 1.93 & -0.174 & 15\\
				& & z$^6$F$_{1/2}$ & 42498.42 & 235.23 & 0.613$\pm$0.01 & 0.671 & 3.60$\pm$0.6 & 4.68 & 4.38 & 0.077 & 17\\
				& & Residual & & & & 0.0092 & & & &\\	\hline	
\end{tabular}
\end{center}
\label{tab:Avalues}
$^a$ Complete description of transition: 3d$^6$($^5$D)4p z$^6$F, z$^6$P, z$^4$P - 3d$^6$($^5$D)4d e$^6$G.\\ Transition energies are from \citet{NJ13} where available or else from \citet{K13}. \\
$^b$ This work\\
$^c$ \citet{K13}\\
$^d$ \citet{RU98web,RU98}\\
$^e$ Uncertainty in the experimental A values\\
$^f$\textit{Residual} is the sum of all theoretical $BF$s for the unobserved lines from this level.
\end{sidewaystable*}%

\end{document}